\documentclass[conference]{IEEEtran}
\usepackage{cite}

\ifCLASSINFOpdf
\usepackage[pdftex]{graphicx}
\else
\fi
\usepackage{amsmath}
\usepackage{amsfonts} 
\usepackage{amsthm}
\usepackage{amssymb}
\usepackage[boxruled]{algorithm2e}
\usepackage{amsthm} 
\usepackage{mathtools} 
\usepackage{textcomp} 

\usepackage{BOONDOX-ds}

\usepackage{bbm}

\newcommand{\pfun}{\mathop{\hbox{$\to$\kern-7pt\raise.9pt\hbox{\scalebox{1}[.55]{$|$}}\kern4pt} }}

\usepackage{array}

\begin{document}

\title{Streaming 1.9 Billion Hypersparse Network Updates per Second with D4M}

\author{\IEEEauthorblockN{Jeremy Kepner$^{1,2,3}$, Vijay Gadepally$^{1,2}$, Lauren
    Milechin$^4$, Siddharth Samsi$^1$, \\ William Arcand$^1$, David  Bestor$^1$, William Bergeron$^1$, 
Chansup Byun$^1$, Matthew Hubbell$^1$, \\ Michael Houle$^1$, Michael Jones$^1$, Anne Klein$^1$, Peter Michaleas$^1$, \\ Julie Mullen$^1$, Andrew Prout$^1$, Antonio Rosa$^1$,  Charles
Yee$^1$, Albert Reuther$^1$
\\
\IEEEauthorblockA{$^1$MIT Lincoln Laboratory Supercomputing Center, $^2$MIT Computer Science \& AI Laboratory, \\ $^3$MIT Mathematics Department, $^4$MIT Department of Earth, Atmospheric and Planetary Sciences}}}
\maketitle

\begin{abstract}
The Dynamic Distributed Dimensional Data Model (D4M) library implements associative arrays in a variety of languages (Python, Julia, and Matlab/Octave) and provides a lightweight in-memory database implementation of hypersparse arrays that are ideal for analyzing many types of network data.  D4M relies on  associative arrays  which combine properties of spreadsheets, databases, matrices, graphs, and networks, while providing rigorous mathematical guarantees, such as linearity.  Streaming updates of D4M associative arrays put enormous pressure on the memory hierarchy.  This work describes the design and performance optimization of an implementation of hierarchical associative arrays that reduces memory pressure and dramatically increases the update rate into an associative array.  The parameters of hierarchical associative arrays rely on controlling the number of entries in each level in the hierarchy before an update is cascaded.  The  parameters are easily tunable to achieve optimal performance for a variety of applications.   Hierarchical arrays achieve over 40,000 updates per second in a single instance.  Scaling to 34,000 instances of hierarchical D4M associative arrays on 1,100 server nodes on the MIT SuperCloud achieved a sustained update rate of 1,900,000,000 updates per second.  This capability allows the MIT SuperCloud to analyze extremely large streaming network data sets.  
\end{abstract}

%
\IEEEpeerreviewmaketitle

\section{Introduction}
\let\thefootnote\relax\footnotetext{This material is based upon work supported by the Assistant Secretary of Defense for Research and Engineering under Air Force Contract No. FA8702-15-D-0001 and National Science Foundation grants DMS-1312831 and CCF-1533644. Any opinions, findings, conclusions or recommendations expressed in this material are those of the author(s) and do not necessarily reflect the views of the Assistant Secretary of Defense for Research and Engineering or the National Science Foundation.}

Networks form the basis of worldwide communication and are estimated to produce many terabytes per second  (TB/ of Internet traffic. The rise of sophisticated cyber threats is increasing rapidly.  Thousands of new malware programs are registered each day, and a majority of all web traffic comes from bots, many of which are malicious in nature.  High performance analysis of network traffic is critical for defending the Internet.  An important technical challenge for analyzing streaming network data is the need to rapidly build and store analyzable network representations of the data \cite{castellana2017high, busato2018hornet,8547563,8547570,8547514,8547572,samsi2017static,kao2017streaming}. In the case of IP network traffic data, there is the additional challenge that the IP address space is much larger than what is observed in a typical data collection, thus the data are hypersparse and require dealing with potentially new endpoint labels at every update.

Development of novel computer network analytics requires: high-level programming environments, massive amounts of network data, and diverse data products for ``at scale'' algorithm pipeline development.  Our team has developed a scalable network analytics platform using the D4M (Dynamic Distributed Dimensional Data Model - d4m.mit.edu) analytics environment and MIT SuperCloud interactive computing environment \cite{kepner2012dynamic,gadepally2018hyperscaling}. D4M combines the power of sparse linear algebra, associative arrays, parallel processing, and distributed databases (such as SciDB and Apache Accumulo) to provide a scalable data and computation system that addresses the big data problems associated with network analytics development. The MIT SuperCloud allows users to interactively process massive amounts of data in minutes on many thousands of cores while using the software and environments most familiar to them.  A key challenge for this pipeline is handling streaming updates of large networks in hypersparse representation.  This paper describes the implementation of a hierarchical approach designed to optimize the performance of the memory hierarchy.  First, associative array mathematics is presented that is the basis of D4M.  Next, is given the design of the hierarchical associative array that leverages these mathematics.  The hierarchical array tuning parameters are described, and results are presented for different parameter settings.  Finally, the performance results of using 34,000 instances on 1,100 compute nodes are presented.

\section{Associative Array Mathematics}

\begin{figure*}[t]
\centering
\includegraphics[width=6in]{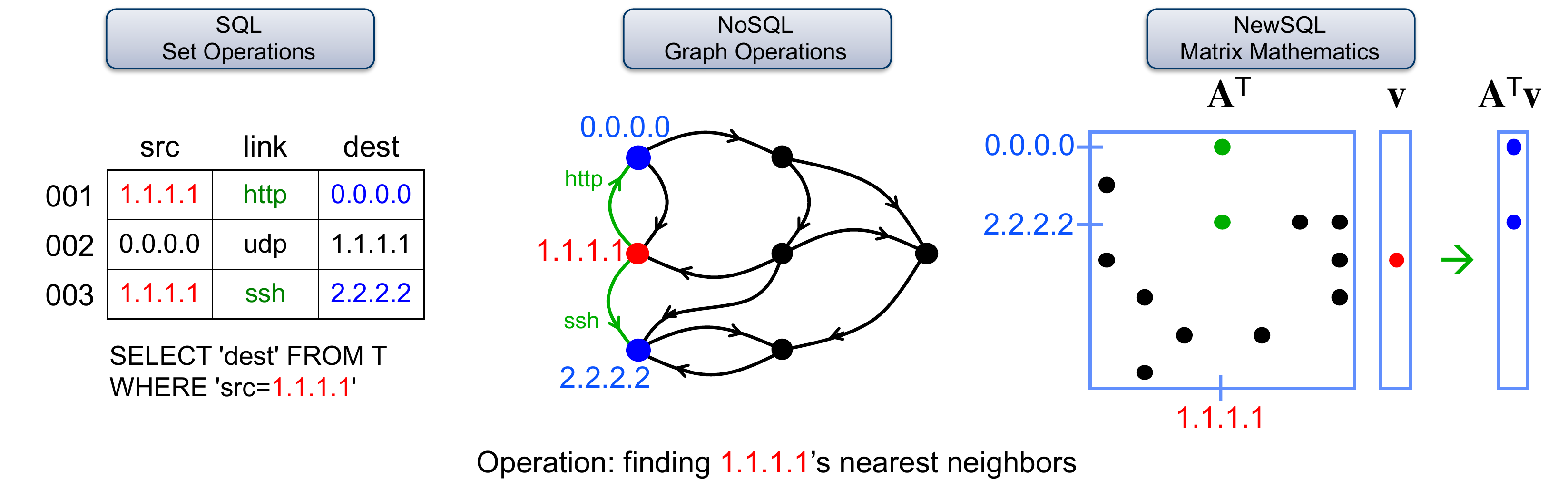}
\caption{Associative arrays combine the properties of databases, graphs, and matrices and provide common mathematics that span SQL, NoSQL, and NewSQL databases, and are ideal for analyzing networks.  The diagram shows the graph operation of finding the neighbors of 1.1.1.1 in each representation.}
\label{fig:AssociativeArrays}
\end{figure*}

Analyzing large-scale networks requires high performance streaming updates of graph representations of these data.  Associative arrays are mathematical objects combining properties of spreadsheets, databases, matrices, and graphs, and are well-suited for representing and analyzing streaming network data (see Fig.~\ref{fig:AssociativeArrays}). 
In many databases, these table operations can be mapped onto well-defined mathematical operations with known mathematical properties.  For example, relational (or SQL) databases \cite{Stonebraker1976,date1989guide,elmasri2010fundamentals} are described by relational algebra \cite{codd1970relational,maier1983theory,Abiteboul1995} that corresponds to the union-intersection semiring ${\cup}.{\cap}$ \cite{jananthan2017polystore}.  Triple-store databases (NoSQL) \cite{DeCandia2007,LakshmanMalik2010,George2011,7322476,Wall2015} and analytic databases (NewSQL) \cite{Stonebraker2005,Kallman2008,Balazinska2009,StonebrakerWeisberg2013,Hutchison2015,gadepally2015graphulo}  follow similar mathematics \cite{kepner2016associative}.  The table operations of these databases are further encompassed by associative array algebra, which brings the beneficial properties of matrix mathematics and sparse linear systems theory, such as closure, commutativity, associativity, and distributivity \cite{kepnerjananthan}.
The aforementioned mathematical properties provide strong correctness and linearity guarantees that are independent of scale and particularly helpful when trying to reason about massively parallel systems.

\begin{figure*}[]
\centering
\includegraphics[width=6in]{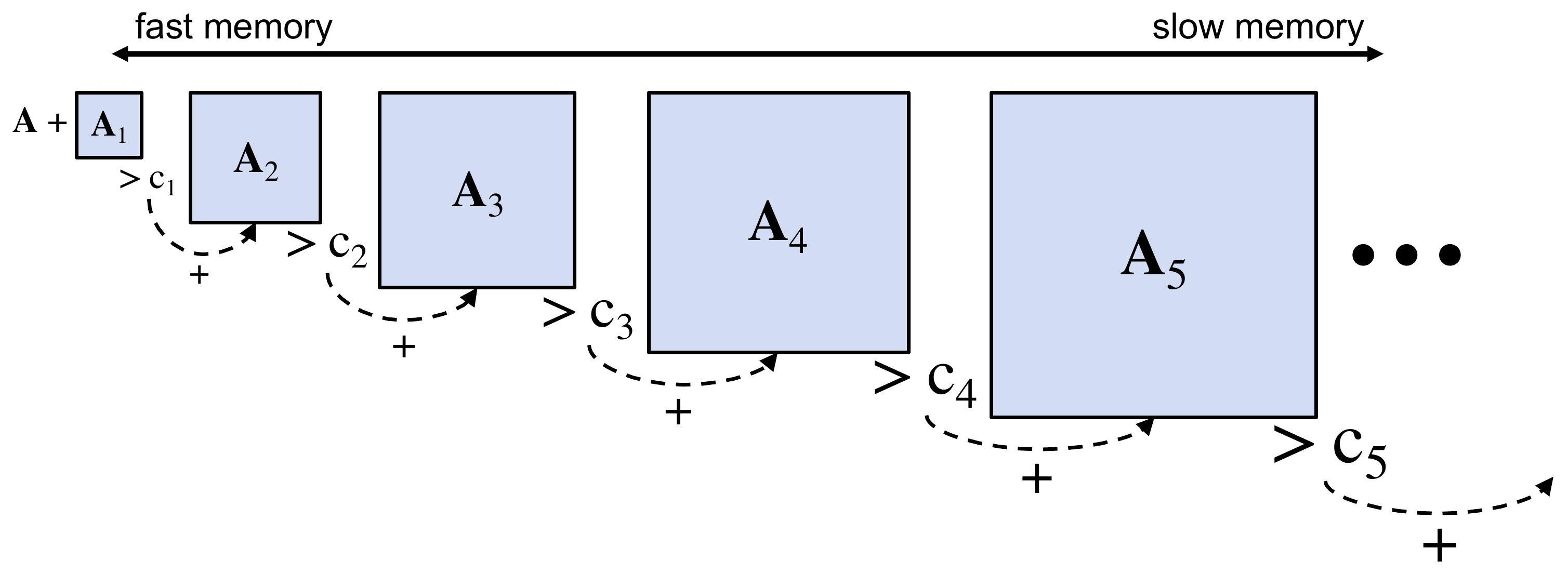}
\caption{Hierarchical associative arrays store increasing numbers of nonzero entries in each layer.  If layer ${\bf A}_i$ surpasses the nonzero threshold $c_i$ it is added to ${\bf A}_{i+1}$ and cleared.  Hierarchical arrays ensure that the majority of updates are performed in fast memory.}
\label{fig:HierarchicalArrays}
\end{figure*}

  The full mathematics of associative arrays and the ways they encompass matrix mathematics and relational algebra are described in the aforementioned references \cite{jananthan2017polystore,kepner2016associative,kepnerjananthan}.  Only the essential mathematical properties of associative arrays are reviewed here.  The essence of associative array algebra is three operations: element-wise addition (database table union), element-wise multiplication (database table intersection), and array multiplication (database table transformation).  In brief, an associative array $\mathbf{A}$ is defined as a mapping from sets of keys to values
$$
  \mathbf{A}: K_1 \times K_2 \to \mathbb{V}
$$
where $K_1$ are the row keys and $K_2$ are the column keys and can be any sortable set, such as integers, real numbers, and strings. The row keys are equivalent to the sequence ID in a relational database table.  The column keys are equivalent to the column names in a database table.  $\mathbb{V}$ is a set of values that forms a semiring $(\mathbb{V},\oplus,\otimes,0,1)$ with addition operation $\oplus$, multiplication operation $\otimes$, additive identity/multiplicative annihilator 0, and multiplicative identity 1. The values can take on many forms, such as numbers, strings, and sets. One of the most powerful features of associative arrays is that addition and multiplication can be a wide variety of operations.  Some of the common combinations of addition and multiplication operations that have proven valuable are standard arithmetic addition and multiplication ${+}.{\times}$, the aforementioned union and intersection ${\cup}.{\cap}$, and various tropical algebras that are important in finance \cite{klemperer2010product,baldwin2016understanding,masontropical} and neural networks \cite{Kepner2017graphblasDNN}: ${\max}.{+}$, ${\min}.{+}$, ${\max}.{\times}$, ${\min}.{\times}$, ${\max}.{\min}$, and ${\min}.{\max}$.

The construction of an associative array is denoted
$$
  \mathbf{A} = \mathbb{A}(\mathbf{k}_1,\mathbf{k}_2,\mathbf{v})
$$
where $\mathbf{k}_1$, $\mathbf{k}_2$, and $\mathbf{v}$ are vectors of the row keys, column keys, and values of the nonzero elements of $\mathbf{A}$.  When the values are 1 and there is only one nonzero entry per row or column, this associative array is denoted
$$
  \mathbb{I}(\mathbf{k}_1,\mathbf{k}_2) = \mathbb{A}(\mathbf{k}_1,\mathbf{k}_2,1)
$$
and when $\mathbb{I}(\mathbf{k}) = \mathbb{I}(\mathbf{k},\mathbf{k})$, this array is referred to as the identity.

Given associative arrays $\mathbf{A}$, $\mathbf{B}$, and $\mathbf{C}$, element-wise addition is denoted
$$
   \mathbf{C} = \mathbf{A} \oplus \mathbf{B}
$$
or more specifically
$$
   \mathbf{C}(k_1,k_2) = \mathbf{A}(k_1,k_2) \oplus \mathbf{B}(k_1,k_2)
$$
where $k_1 \in K_1$ and $k_2 \in K_2$. Similarly, element-wise multiplication is denoted
$$
   \mathbf{C} = \mathbf{A} \otimes \mathbf{B}
$$
or more specifically
$$
   \mathbf{C}(k_1,k_2) = \mathbf{A}(k_1,k_2) \otimes \mathbf{B}(k_1,k_2)
$$
Array multiplication combines addition and multiplication and is written
$$
   \mathbf{C} = \mathbf{A} \mathbf{B} = \mathbf{A} {\oplus}.{\otimes} \mathbf{B}
$$
or more specifically
$$
   \mathbf{C}(k_1,k_2) = \bigoplus_k \mathbf{A}(k_1,k) \otimes \mathbf{B}(k,k_2)
$$
where $k$ corresponds to the column key of $\mathbf{A}$ and the row key of $\mathbf{B}$. Finally, the array transpose is denoted
$$
   \mathbf{A}(k_2,k_1) = \mathbf{A}^{\sf T}(k_1,k_2)
$$
The above operations have been found to enable a wide range of database algorithms and matrix mathematics while also preserving several valuable mathematical properties that ensure the correctness of out-of-order execution.  These properties include commutativity
\begin{eqnarray*}
   \mathbf{A} \oplus \mathbf{B} &=& \mathbf{B} \oplus \mathbf{A} \\
   \mathbf{A} \otimes \mathbf{B} &=& \mathbf{B} \otimes \mathbf{A} \\
   (\mathbf{A} \mathbf{B})^{\sf T} &=& \mathbf{B}^{\sf T} \mathbf{A}^{\sf T}
\end{eqnarray*}
associativity
\begin{eqnarray*}
   (\mathbf{A} \oplus \mathbf{B}) \oplus \mathbf{C} &=& \mathbf{A} \oplus (\mathbf{B} \oplus \mathbf{C}) \\
   (\mathbf{A} \otimes \mathbf{B}) \otimes \mathbf{C} &=& \mathbf{A} \otimes (\mathbf{B} \otimes \mathbf{C}) \\
   (\mathbf{A}  \mathbf{B})  \mathbf{C} &=& \mathbf{A}  (\mathbf{B}  \mathbf{C})
\end{eqnarray*}
distributivity
\begin{eqnarray*}
   \mathbf{A} \otimes (\mathbf{B} \oplus \mathbf{C}) &=& (\mathbf{A} \otimes \mathbf{B}) \oplus (\mathbf{A} \otimes \mathbf{C}) \\
   \mathbf{A} (\mathbf{B} \oplus \mathbf{C}) &=& (\mathbf{A}  \mathbf{B}) \oplus (\mathbf{A}  \mathbf{C}) 
\end{eqnarray*}
and the additive and multiplicative identities
$$
   \mathbf{A} \oplus \mathbs{0} = \mathbf{A} ~~~~~~~~~~~~ \mathbf{A} \otimes \mathbs{1} = \mathbf{A} ~~~~~~~~~~~~ \mathbf{A} \mathbb{I} = \mathbf{A}
$$
where $\mathbs{0}$ is an array of all 0, $\mathbs{1}$ is an array of all 1, and $\mathbb{I}$ is an array with 1 along its diagonal.  Furthermore, these arrays possess a multiplicative annihilator
$$
   \mathbf{A} \otimes \mathbs{0} = \mathbs{0} ~~~~~~~~~~~~ \mathbf{A} \mathbs{0} = \mathbs{0}
$$
Most significantly, the properties of associative arrays are determined by the properties of the value set $\mathbb{V}$.  In other words, if $\mathbb{V}$ is linear (distributive), then so are the corresponding associative arrays.

Intersection $\cap$ distributing over union $\cup$ is essential to database query planning and parallel query execution over partioned/sharded database tables \cite{booth1976distributed,shaw1980relational,stonebraker1986case,barroso2003web,curino2010schism,pavlo2012skew,corbett2013spanner}.  Similarly, matrix multiplication distributing over matrix addition ensures the correctness of massively parallel implementations on the world's largest supercomputers \cite{dongarra2003linpack} and machine learning systems \cite{moller1993scaled,werbos1994roots,chetlur2014cudnn}.  In software engineering, the scalable commutativity rule guarantees the existence of a conflict-free (parallel) implementation \cite{clements2015scalable,clements2017scalable,bhat2017designing}.  The full mathematics of associative arrays and the ways they encompass matrix mathematics and relational algebra are described in the aforementioned references \cite{jananthan2017polystore,kepner2016associative,kepnerjananthan}.

\section{Hierarchical Associative Arrays}

The D4M library implements associative arrays in a variety of languages (Python, Julia, and Matlab/Octave) and provides a lightweight in-memory database.  Most implementations use sorted strings for the row, column, and value labels, and standard sparse matrix to connect the triples. Associative arrays are designed for block updates. Streaming updates to a large associative array requires a hierarchical implementation to optimize the performance of the memory hierarchy (see Fig.~\ref{fig:HierarchicalArrays}).   Rapid updates are performed on the smallest arrays in the fastest memory.  The strong mathematical properties of associative arrays allow a hierarchical implementation of associative arrays to be implemented via simple addition. All creation and organization of hypersparse row and column labels are handled naturally by the associative array mathematics.  If the number of nonzero (nnz) entries exceeds the threshold $c_i$, then ${\bf A}_i$ is added to ${\bf A}_{i+1}$ and ${\bf A}_i$ is cleared.  The overall usage is as follows
\begin{itemize}
\item Initialize $N$-layer hierarchical array with cuts $c_i$
\item Update by adding data ${\bf A}$ to lowest layer
$$
  {\bf A}_1 = {\bf A}_1 + {\bf A}
$$
\item If ${\rm nnz}({\bf A}_1) > c_1$, then
$$
  {\bf A}_2 = {\bf A}_2 + {\bf A}_1
$$
and clear ${\bf A}_1$.
\end{itemize}

\noindent The above steps are repeated until ${\rm nnz}({\bf A}_i) \leq c_i$ or $i = N$.  To complete all pending updates for analysis, all the layers are added together
$$
   {\bf A} = \sum_{i=1}^{N} {\bf A}_i
$$
Hierarchical arrays dramatically reduce the number of updates to slow memory.  Upon query, all layers in the hierarchy are summed into the largest array.  The cut values $c_i$ can be selected so as to optimize the performance with respect to particular applications.  The majority of the complex updating is performed by using the existing D4M associative array addition operation. The corresponding Matlab/Octave D4M code for performing the update is direct translation of the above mathematics as follows

\noindent \rule{\columnwidth}{0.5pt}

{\tt
\noindent function Ai = HierAdd(Ai,A,c);

   Ai\{1\} = Ai\{1\} + A;
   
   for i=1:length(c)
   
   ~  if (nnz(Ai\{i\}) > c(i))
     
   ~~~    Ai\{i+1\} = Ai\{i+1\} + Ai\{i\};
       
   ~~~    Ai\{i\} = Assoc(\textquotesingle\textquotesingle,\textquotesingle\textquotesingle,\textquotesingle\textquotesingle);
       
   ~  end
     
   end

\noindent end 
}

\noindent \rule{\columnwidth}{0.5pt}

\section{Performance Optimization}

\begin{figure}[]
\centering
\includegraphics[width=\columnwidth]{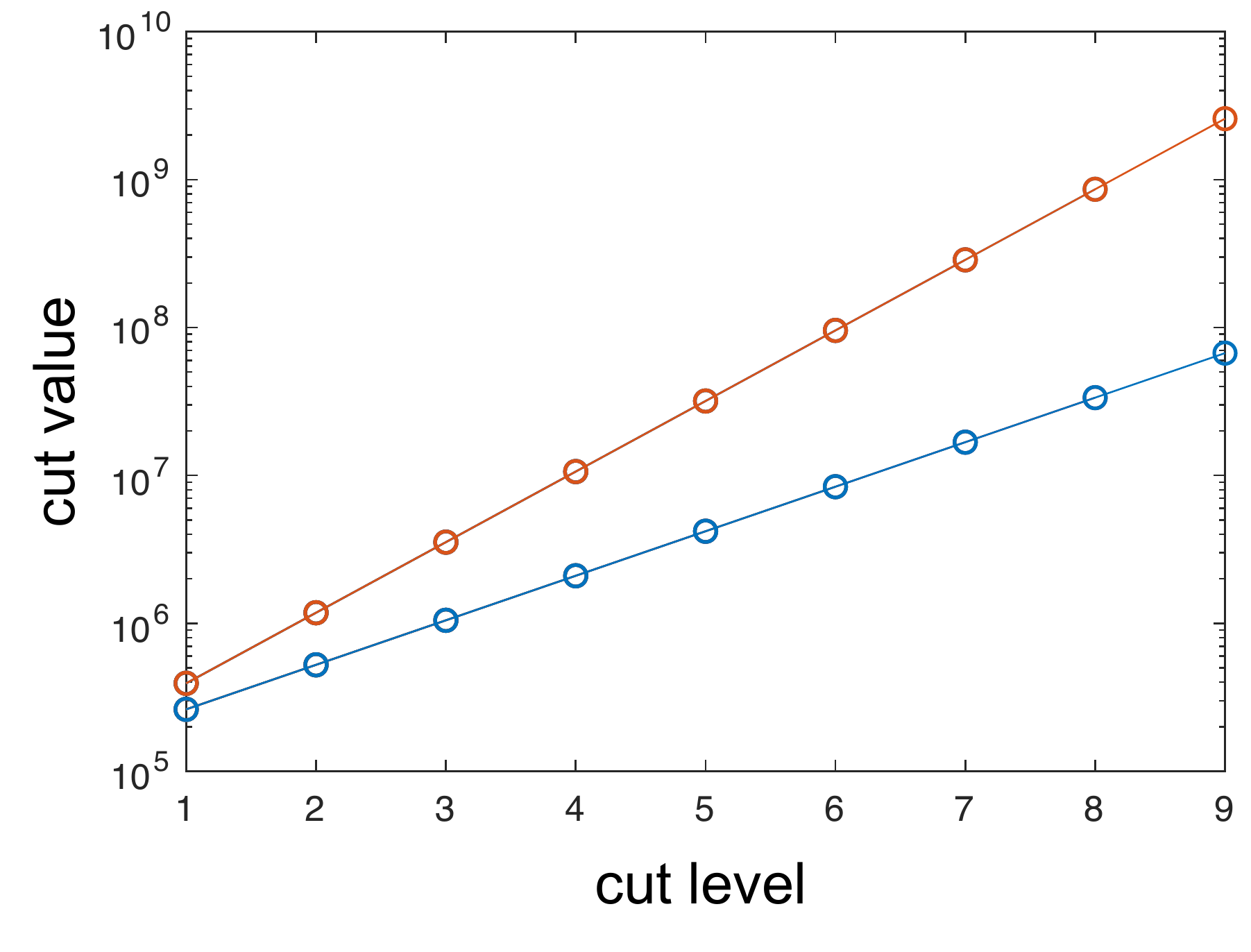}
\caption{Number of nonzeros (nnz) and cut value. Cut values determine the size and update pattern of the hierarchical array. Few widely spaced cuts require less memory, but minimize benefit.  Many closely spaced cuts maximize benefit, but require more memory
}
\label{fig:CutValue}
\end{figure}

\begin{figure}[]
\centering
\includegraphics[width=\columnwidth]{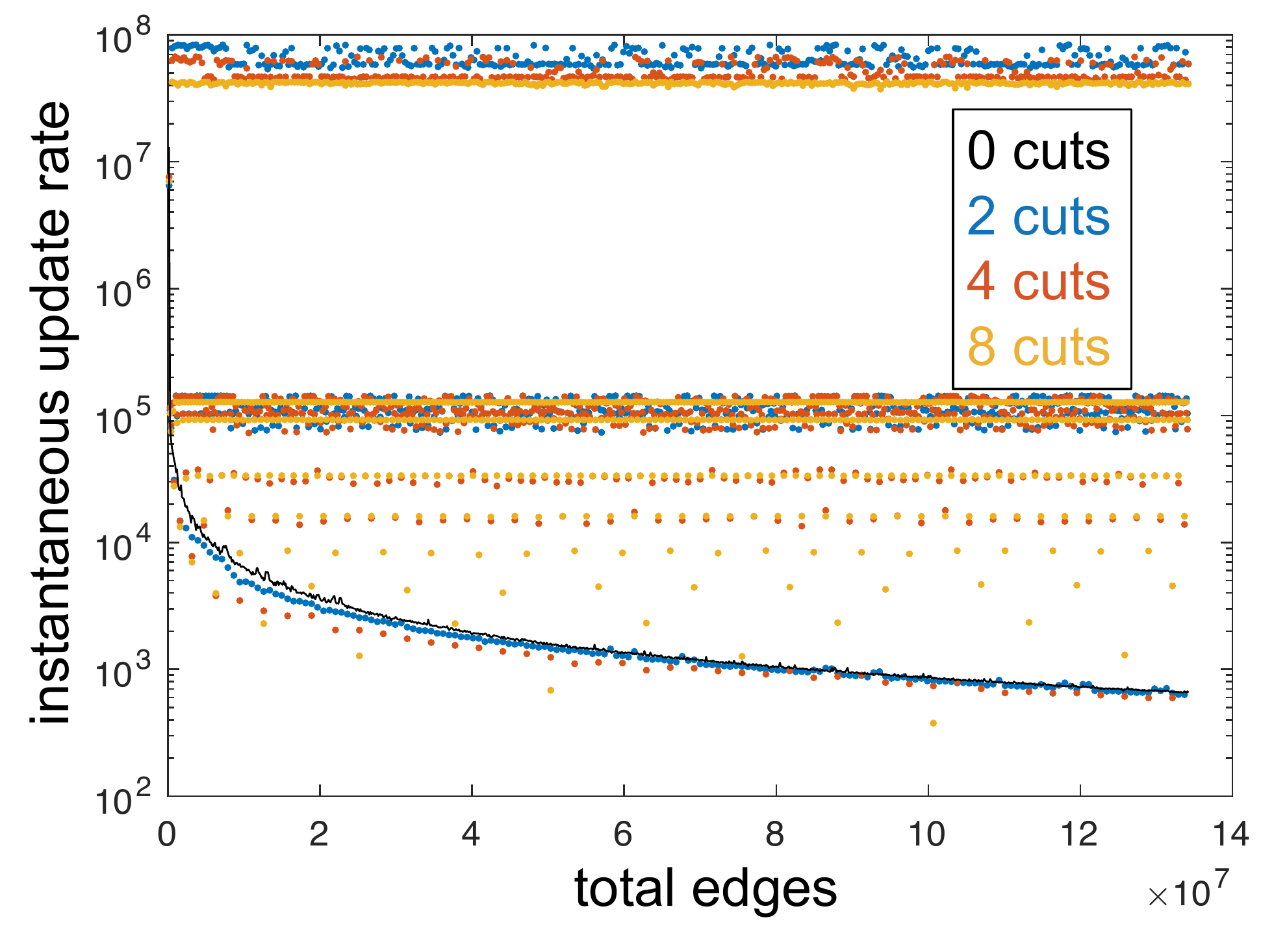}
\caption{Instantaneous update rate on a single core. 0 cuts results in steadily decreasing performance as the total edges in the graph increase. 2, 4, 8 cuts reduce the number of updates to slow memory and increase the update rate.}
\label{fig:UpdateRateVsEdges}
\end{figure}

\begin{figure}[]
\centering
\includegraphics[width=\columnwidth]{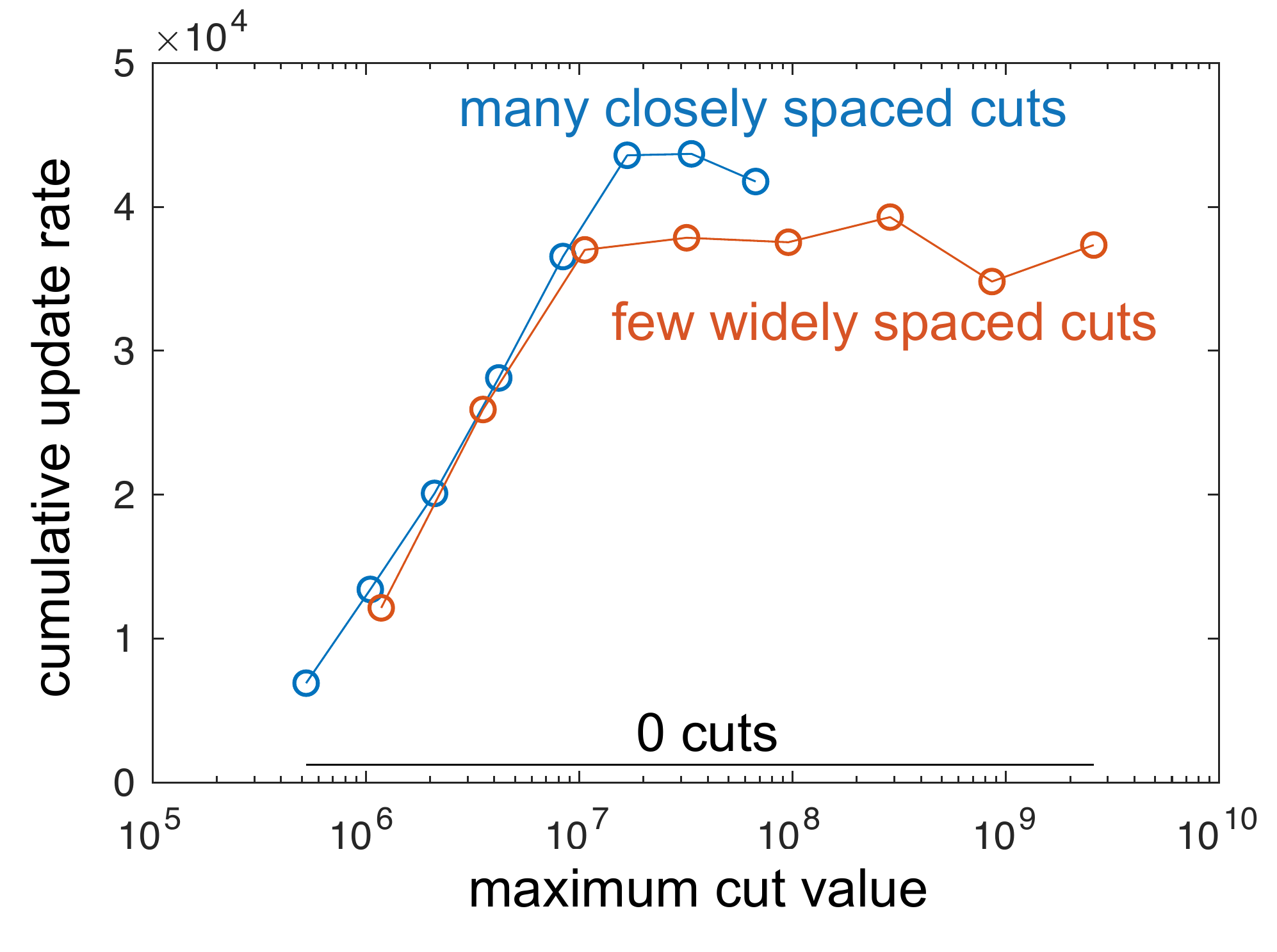}
\caption{Cumulative update rate (single core). Many closely spaced cuts produces highest update rate.  Both spacings offer significant performance advantage over 0 cuts.}
\label{fig:UpdateRateSingleCore}
\end{figure}

The performance of a hierarchical associative array for any particular problem is determined by the number of layers $N$ and the cut values $c_i$.  The parameters are tuned to achieve optimal performance for a given problem.  Examples of two sets of cut values with different $c_1$  and different ratios between cut values are shown in Figure~\ref{fig:CutValue}.  These sets of cut values allow exploration of the update performance of many closely spaced cuts versus few widely spaced cuts.  Figure~\ref{fig:UpdateRateVsEdges} shows the single instance (single processor core) performance for different numbers of layers and cut values on a simulated Graph500.org R-Mat power-law network data. The data set contains 100,000,000 connections that are inserted in groups of 100,000.  In general, the hierarchical performance is much better than the non-hierarchical implementation (0-cuts) and increases until the last cut is above the total number of entries in the data.   The instantaneous insert rate of each group of network connections is shown in Figure~\ref{fig:UpdateRateSingleCore} and indicates that more cuts allow more updates to occur quickly because they are taking place in faster memory.

\section{Scalability Results}

\begin{figure}[]
\centering
\includegraphics[width=\columnwidth]{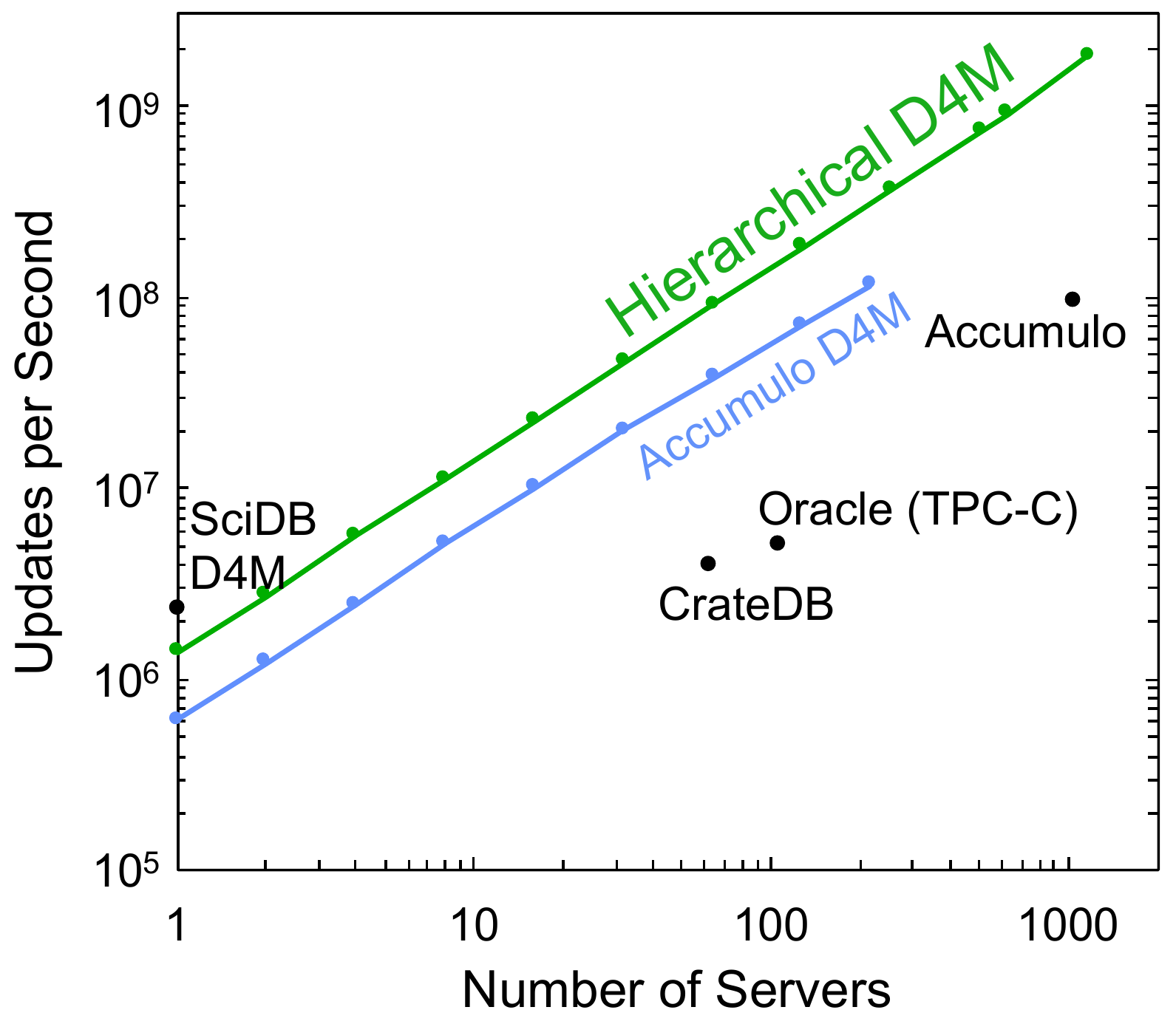}
\caption{Update rate as a function of number of servers for hierarchical D4M associative arrays and other previous published work: Accumulo D4M \cite{kepner2014achieving}, SciDB D4M \cite{samsi2016benchmarking}, Accumulo \cite{sen2013benchmarking}, Oracle TPC-C benchmark, and CrateDB \cite{CrateDB}}
\label{fig:UpdateRate}
\end{figure}

  The scalability of the hierarchical associative arrays are tested using a power-law graph of 100,000,000 entries divided up into 1,000 sets of 100,000 entries.  These data were then simultaneously loaded and updated using a varying number of processes on varying number of nodes on the MIT SuperCloud up to 1,100 servers with 34,000 processors. This experiment mimics thousands of processors, each creating many different graphs of 100,000,000 edges each.  In a real analysis application, each process would also compute various network statistics on each of the streams as they are updated.   The update rate as a function of number of server nodes is shown on Fig.~\ref{fig:UpdateRate}.  The achieved update rate of 1,900,000,000 updates per second is significantly larger than the rate in prior published results. This capability allows the MIT SuperCloud to analyze extremely large streaming network data sets.

\section{Conclusion}

The D4M implementation of associative arrays provides a lightweight in-memory database ideal for analyzing  hypersparse network data.   Associative array mathematics combines properties of spreadsheets, databases, matrices, graphs, and networks and provides strong linearity guarantees.  Streaming data into associative arrays puts enormous pressure on a memory hierarchy.  D4M hierarchical associative arrays reduce memory pressure and increase update performance.  The linearity properties of associative arrays allow a hierarchical associative array to be implemented using simple addition operations.  The performance of hierarchical associative arrays comes from controlling the number entries at each level and can be tuned for any particular application. Hierarchical arrays achieve over 40,000 updates per second in a single instance and are significantly faster than non-hierarchical associative arrays.  Scaling to 34,000 instances of hierarchical D4M associative arrays on 1,100 server nodes on the MIT SuperCloud achieved a sustained update rate of 1,900,000,000 updates per second. 

\section*{Acknowledgement}

The authors wish to acknowledge the following individuals for their contributions and support: Bob Bond, Alan Edelman, Charles Leiserson, Dave Martinez, Mimi McClure, Victor Roytburd, Michael Wright.


%
%




\bibliographystyle{ieeetr}

\bibliography{aarabib}
%

\end{document}